# Assert Use and Defectiveness in Industrial Code

S. Counsell, T. Hall
*Dept. of Computer Science*
*Brunel University*
*London, UK*
steve.counsell@brunel.ac.uk

T. Shippey, D. Bowes
*Dept. of Computer Science*
*University of Hertfordshire*
*Hertfordshire, UK*
t.shippey@herts.ac.uk

A. Tahir
*School of Eng. & Advanced*
*Tech. Massey University*
*New Zealand*
a.tahir@massey.ac.nz

S. MacDonell
*Dept. of Inf. Science*
*University of Otago*
*New Zealand*
stephen.macdonell@otago.ac.nz

**Abstract**

*The use of asserts in code has received increasing attention in the software engineering community in the past few years, even though it has been a recognized programming construct for many decades. A previous empirical study by Casalnuovo showed that methods containing asserts had fewer defects than those that did not. In this paper, we analyze the test classes of two industrial telecom Java systems to lend support to, or refute that finding. We also analyze the physical position of asserts in methods to determine if there is a relationship between assert placement and method defect-proneness. Finally, we explore the role of test method size and the relationship it has with asserts. In terms of the previous study by Casalnuovo, we found only limited evidence to support the earlier results. We did however find that defective methods with one assert tended to be located at significantly lower levels of the class position-wise than non-defective methods. Finally, method size seemed to correlate strongly with asserts, but surprisingly less so when we excluded methods with just one assert. The work described highlights the need for more studies into this aspect of code, one which has strong links with code comprehension.*

## 1. INTRODUCTION AND MOTIVATION

Asserts are widely acknowledged as a powerful automated tool for detecting and localizing faults in programs [3]. They are used as a checking and verification mechanism for what the program "should do" and for assisting during the debugging, deployment and testing stages of development. Using asserts can also improve the reliability of the program, since they provide a means of systematic error-checking and making explicit what code is trying to do. A number of recent studies [1, 4, 5, 9] have explored the role of asserts in code. The work described in this paper is motivated by previous work of Casalnuovo et al., [1]. In that study, the use of asserts in a large set of C and C++ projects on GitHub were studied. Results showed that programs using asserts had fewer defects compared to programs with no asserts. The study underlined the important role that asserts played in the quality of software [1]. These findings were in-line with the previous results of Kudrjavets et al., [9]. In this paper, we also investigate the issue of asserts and defect-proneness in test classes. We explore the physical positioning of asserts and whether test method size has a role in their number. In terms of reflecting the previous study by Casalnuovo, only limited evidence to support earlier results was found. We did, however, find evidence that defective methods with one assert tended to be located at lower levels of the class "position-wise"; Finally, we found a positive, significant relationship between number of asserts and method size, but to varying degrees depending on whether the method was defective or not. The remainder of the paper is organized as follows. In the next section, we describe related work and the data collected. In Section III, we analyze the data collected on four themes before discussing the results and the threats to validity (Section IV). Finally, we conclude and point to future work in Section V.

## 2. RELATEDWORK/PRELIMINARIES

### A. Related work

Asserts in programming languages have been the subject of significant previous interest especially in the areas of error-checking (e.g., [12]) and program verification ([e.g., [2]). The idea of using asserts as a means of program verification can be traced back to Floyd [7]. Later, Yau and Cheung [15] used asserts for automatic run-time checking. The use of asserts has been advocated by many authors and researchers; McConnell [10] advises developers to implement the use of asserts in their programming practices to promote automatic checking for program failures. Fowler et al., [6] suggests 'introduce assertion' as one the 72 refactorings when assumptions need to be made explicit in the code. Meyer in "applying design by contract" [11] advocated the use of asserts as a correction methodology and for establishing contract pre- and post- conditions. The approach of contracts and asserts was implemented in the Eiffel programming language and the Turing language was one of the earliest languages to support the use of asserts [8]. The history of asserts usage in programming languages was detailed by Clarke and Rosenblum [3] and several studies have investigated the use of asserts (or more generally the use of contracts) in both open- and closed-source software. The work of Kudrjavets et al., [9], based on two Microsoft projects, showed that the density of bugs decreased when the density of asserts increased. Estler et al., [5] studied the use of pre- and post-conditions in twenty-one OO projects (written in Java, C# and Eiffel). The study found that the percentage of program elements that included contracts/asserts was above 33% for most projects and tended to be stable over time. Clearly, some insightful work has been done in the area of asserts. However, our knowledge of their characteristics and influence is still largely unknown and leaves many open research questions.

**B. Preliminaries**
The two systems used in the study were written using an agile approach and had been in production for several years. Pair programming, TDD and daily stand-ups were all features of the development practice at the company, based in London. In terms of data collected, the number of asserts and lines of code (LOC) in each of the methods was extracted using the JHawk tool [14]. To identify defective methods, we used the SZZ approach since it has been used in many previous studies [13]. SZZ is a fault linking algorithm described by Sliwerski et al., [13] and matches a fault fix described in a defect tracking system with the corresponding commit in a version control system that 'removed' the defect. By backtracking through the version control records, it is possible to identify earlier code changes which ended up being 'fixed'. It is assumed that the earlier code changes inserted the defect. The module of code (in our case a method) is labelled as defective between the time the defect was inserted and the time it was fixed. Using this technique it is possible to identify, for a particular snapshot of the code base, which modules (methods) are defective and which are not. Finally, we note that for each method we indicate whether that method is defective on a binary (yes/no) basis. This is in contrast to collecting number of defects in each method.

## 3. DATA ANALYSIS

In the next sections, we examine asserts in the two systems from four perspectives. Firstly, for the propensity of methods with asserts to be defective; secondly, a comparison of methods with one assert *versus* those with more than one (in terms of defect-proneness). Thirdly, the role that the position of a method containing asserts plays in defect-proneness and, finally, the size of method and its relationship with asserts.

**A. Defect propensity (asserts vs. methods)**
We explore firstly whether test methods containing at least one assert were less likely to be defective (than those that contained no asserts). This is a partial replication of the work by Casalnuovo [1] which showed that methods containing asserts had fewer defects than those that did not. For System one, 1,232 methods of the 10,504 (11.73%) contained at least one assert (9,272 methods therefore contained zero asserts). Table 1 shows the distribution of those 1,232 asserts in five separate numerical intervals. So, for example, 23 methods contained between 10 and 19 assert statements (inclusive); methods with a single assert numbered 788.

Table 1. Distribution of asserts (System one)

| >=20 <= 24 | >=10 <=19 | >=5<=9 | >=2<=4 | =1 | Total |
|---|---|---|---|---|---|
| 8 | 23 | 64 | 349 | 788 | 1,232 |

Of the 9272 methods in System one with zero asserts, 810 were defective (8.74%). In the five categories shown in Table 1 from a total of 1,232 methods, 107 were defective, representing 8.69%. Comparison of these values does not support the view that methods with asserts are less defective than those with asserts; an almost identical proportion of each category is defective. Carrying out the test for defective methods (zero asserts versus one or more asserts) returned a Z value of - 30.20, significant at the 1% level; For System two, 1,589 of the 12,038 (13.20%) methods in total contained at least one assert (10,449 methods did not therefore contain a single assert). Table 2 shows the distribution of asserts for System two in the same format as Table 1. Of the 10,449 methods that did not include an assert statement, 1,173 were defective (11.23%). Of the 1,589 methods with at least one asset, 219 were defective (13.78%).

For System two, there is, again, only limited support for the claim that methods with at least one assert are more defect-prone than those without asserts - the difference between the two groups is less than 3%. Overall, we cannot therefore conclude that methods containing asserts were and more or less defective than methods without asserts, for the two systems studied.

Table 2. Distribution of asserts (System two)

| >=20 <= 37 | >=10 <=19 | >=5<=9 | >=2<=4 | =1 | Total |
|---|---|---|---|---|---|
| 4 | 7 | 55 | 326 | 1,197 | 1,589 |

**B. One assert versus many**
The values in Tables 1 and 2 show that the vast majority of methods contained just a single assert. For System one, 85.47% of all asserts were single asserts in the method. For System two, the corresponding value was 75.33%. One question that arises from the preceding analysis is whether a single assert in a method is as "effective" as one with many asserts in a method (i.e., >1). In other words, does the number of asserts in a method make any difference to the likelihood of that method being defective? We therefore analyzed the data to determine if the defect profile of the former set of methods (i.e., those in column 5 in Tables 1 and 2) was different to that of methods with more than one assert (obtained by summing columns 1-4 in Tables 1 and 2). For System one, of the 788 methods with a single assert, 59 were defective (7.49%). Of the 444 remaining methods (in columns 1-4 of Table 1), 48 were defective (10.81%). For System 2, of the 1197 with a single assert, 171 were defective (14.29%). For all remaining methods with more than one assert (392), the number of defective methods was also 48 (12.24%). These results therefore suggest that for the two systems, there is no evidence that a method containing more than one assert is any less (or more) defective than one with just a single assert. We do not find support for the earlier work of Casalnuovo [1]. This is an interesting observation from the perspective of assertion use generally. Casalnuovo states that: "*The effect of asserts on bugs in the count model is almost insignificant, and the* *magnitude of the effect is negligible overall. Both models together indicate that adding the first assert to a file has a significant and sizable effect on bugs, but after the first, on average for all developers, adding additional asserts has no appreciable difference*". We accept that we study asserts on a binary (yes/no) defect basis where the actual number of defects may have been more useful in this instance; however, we feel that the result is still revealing and provides an insight into the relative effectiveness of asserts.

### C. Position of an assert

One issue which may be relevant to our understanding of assert usage is whether the physical placement of asserts in a class (i.e., its position in a class) is related to method defectiveness or not. Table 3 shows the data of start line of a method (i.e., where the method starts physically in the class) for all methods (for both Systems one and two) where there was just one assert *and* where the method was defective or non-defective. For example, in System one, the mean start line of a method with one assert in it and which was defective was 139.56 (with median 111); 59 methods fell into this category. We also include the standard deviation values (SD). For non-defective methods with at least one assert, the corresponding mean start line was 79.52 (median 52). In other words, defective methods with asserts are placed far lower down in the class than methods with asserts and which are non-defective in terms of physical lines of code.

Table 3. Start-line analysis of asserts

| Category | Mean | SD | Median | # Methods |
|---|---|---|---|---|
| System one ||||| 
| Defective | 139.56 | 84.55 | 111 | 59 |
| Non-defective | 79.25 | 94.36 | 52 | 729 |
| System two ||||| 
| Defective | 147.04 | 53.55 | 68 | 171 |
| Non-defective | 62.79 | 183.67 | 47 | 1,026 |

The Mann Whitney U test is a non-parametric test and determines the likelihood of a value taken from one sample being the same as or greater than a value taken from a different independent sample. Carrying out a Mann Whitney U test (defective versus non-defective for methods with a single assert) gave a Z value of -6.53 (significant at the 1% (0.01) level) for System one and a Z value of -5.45 for System two, also significant at the 1% level. We conclude that there is a negative significant difference in the positions in a class between defective and non-defective methods (in the case where methods had a single assert). To try and understand this result, we asked one of the lead developers at the company about this effect and why it was that methods at the end of the class would be more defective. One explanation offered by the developer was that new methods were usually added at the end of the class during maintenance (and in response to new functionality being added to the system), as we might expect. The systems they maintained, however, were becoming more defective over time and requirements were getting harder to implement correctly (because of issues such as technical debt and changing team members). More recent tests at the end of a file were therefore more likely to be defective than the older tests at the top of the class and this might go some way to explaining our result.

### D. Asserts and size

The final aspect of asserts that we explore is the extent to which asserts are related to method size. The study by Estler et al., using Java, Eiffel and C# systems found that the number of asserts correlated positively with project size. Figure 1 shows the correlation between the numbers of asserts *and* the size of methods for System one excluding all methods where there were zero asserts. The largest number of asserts in any single method for this systems was 24.

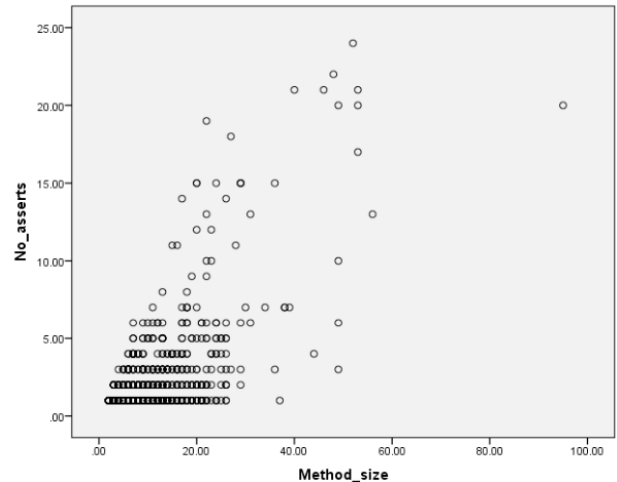

Figure 1. System one correlation (no. of asserts vs. method size)

The correlation values for all asserts were all found to be significant at the 1% level (Kendall's (0.41) Spearman's rank correlation coefficient (0.50) and Pearson's (0.67)). Both Kendall's and Spearman's correlations are non-parametric tests and make no assumption about the normality of the data. Pearson's, on the other hand, assumes that the data is normally distributed (we include all three for the sake of completeness). Interestingly, if we then decompose the data into defective and non-defective, these correlation values change to just 0.42 (Kendall's), 0.51 (Spearman's) and 0.68 (Pearson's) for the set of defective methods and 0.31 (Kendall's), 0.38 (Spearman's) and 0.51 (Pearson's) for the non-defective methods, a far weaker set of correlation values. This implies that there may well be a positive effect of using *more* asserts in identification of defective methods. Figure 2 shows the corresponding scatter plot for System two (method size versus number of asserts). The largest number of asserts in any single method for this system was 37 (with 234 LOC).

The correlation values for this scatter plot are 0.34 (Kendall's), 0.39 Spearman's) and 0.62 (Pearson's), all of which are significant at the 1% level. If we then decompose the data into defective and non-defective categories, we get correlation values of 0.34 (Kendall's), 0.41 (Spearman's) and 0.46 (Pearson's) for the set of defective methods, all significant at the 1% level and 0.34, 0.39, and 0.63 for the set of non- defective methods, all of which are significant at the 1% level. There is a slightly weaker set of results for the non-defective methods if we consider the non-parametric correlation values, again supporting the view that the more asserts the better. We did not therefore find overwhelming evidence in support of the view that asserts are directly related to method size.

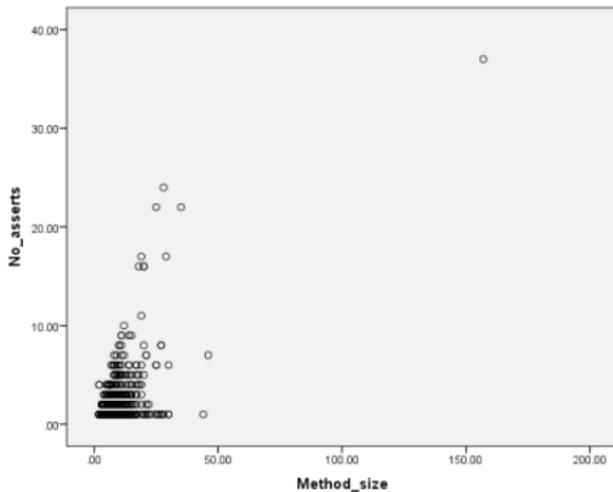

Figure 2. System two correlation (no. of asserts vs. method size)

## 4. STUDY IMPLICATIONS AND VALIDITY THREATS

The preceding analysis raises the question as to the implications for the tester/developer. The weight of evidence does seem to point to methods with just a single assert being different in their link with defects than methods with many asserts. In some ways, it may be the case that a carefully chosen single assert is as useful and effective as a methods liberally spread with asserts. This is the view that Casalnuovo et al., expressed when emphasizing the value of the first assert and is an issue that needs further study. We also have to consider the result of assertion positioning and the view of the developer. Clearly, evolution is a factor in the deterioration of test code just as it for production code. The result suggests that more refactoring of test code and reversal of code decay might prove worthwhile. An interesting further study would be to examine the patterns in the addition of asserts over time. Finally, the link between asserts and size was interesting – it again suggests that a single assert has a stronger correlation with size than if we include methods with multiple asserts. The analysis in the paper raises a number of threats to validity. Firstly, we only studied two systems from an industrial partner and we cannot therefore easily extend the results to other industrial or open-source systems. Secondly, the two systems we used in this study were both telecoms systems; this might restrict the extent to which we could generalize our conclusions to other application domains. Thirdly, the number of asserts in the methods of the systems was quite low (between 10% and 14%) – so our sample sizes were quite low taking the 'wider' picture. However, this might have been expected: the company reported very few low-level coding defects; many of the problems they faced were less to do with unit testing and more due to the interfaces with supplier systems. Finally, we only flagged a method as defective or non-defective without including the number of defects. However, the original purpose of the study was to try and link asserts with a coarse view of defect-proneness.

## 5. CONCLUSIONS AND FUTUREWORK

In this paper, we explored the use of asserts in two industrial telecoms systems. We explored four aspects of asserts. Firstly, for the propensity for methods with asserts to be defective – *we found no evidence that methods containing asserts were more defective than methods without for the two systems studied*. Secondly, we compared methods with one assert *versus* those with more than one (in terms of defect-proneness). *We found no evidence that a method containing more than one assert was any less (or more) defective than one with just a single assert*. Thirdly, the role that the position of a method containing asserts plays in defect-proneness. *We found a statistically significant difference in the positions in a class between defective and non-defective methods where methods had a single assert)*. Finally, the size of method and its relationship with asserts. *No overwhelming support for a link between asserts and method size was found.*

All the analysis in the paper has been made at the method level, rather than the class level. Most studies in the past have looked at systems at the latter. We see a lot of research scope for looking at systems at this lower level of abstraction rather than at higher levels for the detail it offers, especially when classes can be very large. In terms of future work, we will explore more systems both industrial/open-source (the company has ~100 similar systems and defects. One aspect we haven't explored in this paper is test quality versus quantity and whether an optimum level of asserts exists beyond which there are diminishing returns; this is also a subject of further work.

## ACKNOWLEDGEMENTS

This work was funded by the UK's Engineering and Physical Sciences Research Council (EPSRC) Grant EP/L011751/1.